\documentclass[epj]{svjour}
\usepackage{graphicx,bm}
\begin{document}
\title{Exploration of hyperfine interaction between constituent
quarks via $\eta$ productions}
\author{Jun He\inst{1,2} \and S.~G.~Yuan\inst{1,3}
\and H.~S.~Xu\inst{1}}
\institute{{Nuclear theory group, Institute of Modern Physics, Chinese Academy of Sciences,
Lanzhou 730000, China}\and
{Research Center for Hadron and CSR Physics,
Institute of Modern Physics of CAS and Lanzhou University, Lanzhou
730000, China}\and
{Graduate University of Chinese Academy of
Sciences, Beijing 100049, China}
}
\date{Received: date / Revised version: date}
\abstract{
In this work, the different exchange freedom, one gloun, one pion or
Goldstone boson, in constituent quark model is investigated, which is
responsible to the hyperfine interaction between constituent quarks,
$via$ the combined analysis of the $\eta$ production processes,
$\pi^{-}p\rightarrow\eta n$ and $\gamma p\rightarrow\eta p$. With the
Goldstone-boson exchange, as well as the one-gluon or
one-pion exchange, both the spectrum and observables, such as, the
differential cross section and polarized beam asymmetry, are fitted to
the suggested values of Particle Data Group and the experimental data.
The first two types of exchange freedoms give acceptable description
of the spectrum and observables while the one pion exchange can not
describe the observables and spectrum simultaneously, so can be
excluded. The experimental data for the two processes considered here
strongly support the mixing angles for two lowest $S_{11}$ sates and
$D_{13}$ states as about $-30^\circ$ and $6^\circ$ respectively.
\PACS{
{12.39.Pn}{Potential models }\and
{13.60.Le}{Meson production}\and 
{14.20.Gk}{Baryon resonances (S=C=B=0)} 
} 
% end of PACS codes
} %end of abstract
\maketitle
\section{Introduction}\label{sec:1}

How to understand the baryon spectrum in the nonperturbative QCD
dynamics is still a wide open sector of particle physics after the
discovery of the first baryon resonance $\Delta$ by Fermi in the
fifties of last century. Due to the difficulty of dealing with the QCD
in the nonperturbative energy region, many phenomenological models are
proposed to describe  the internal structure of the hadron.  Among the
phenomenological models the constituent quark model (CQM) achieved a
vast of successes and become the basis of discussion about the hadron
spectrum to some extent, such as the ``missing resonances''. Even some
assumptions of CQM have been confirmed by the modern lattice QCD
calculation, such as the massive $u$ $d$  quark with mass about 300
MeV~\cite{Bowman:2005vx}. The relation between CQM and large
$1/N_c$ approach and QCD sum rule is also
investigated~\cite{Melikhov:2004uk,Pirjol2009,Pirjol:2010th}, which
supports CQM as a right effective description in nonperturbative energy
region.

Expect the confinement potential an important ingredient of CQM is the
hyperfine interaction between constituent quarks, which is related to
the mass splitting of the states in the same multiplet, such as the
two negative parity $S_{11}$ nucleon resonances. A popular hyperfine
interaction is from the one-gluon exchange (OGE) inspired by
the fundamental theory of strong interaction QCD
\cite{Isgur1978a,Isgur1979,Capstick1986}.  It is also called as
Isgur-Karl model due to the surprising success of their version of OGE
in the description of the baryon spectrum and many properties of
hadron and the corresponding
resonances\cite{Koniuk1980,Capstick2000a}. A modern model, the
Goldstone-boson exchange (GBE) model based on the spontaneous breaking
of chiral symmetry of QCD \cite{Melde2007,Choi2010}, which
is extended from an early version, one-pion exchange (OPE) model
\cite{Glozman1996} proposed by Riska and Glozman about 2000 , can also
be applied to describes the baryon spectrum and predict decay
properties of resonances \cite{Melde2007} and axial charges
of baryon \cite{Choi2010}. Besides, Manohar and Georgi argued that
both Goldstone boson and gluon effective degrees of freedom survive in
the spontaneously broken chiral symmetry and confinement energy
scales~\cite{Manohar:1983md,Collins:1998ny}.

A problem arises: which one of them is right? It is natural to judge
the different hyperfine interactions in the basic theory of strong
interaction QCD.  However, the difficulty of application of QCD in the
low energy nonpertubative region is just the reason why we adopt the
phenomenological model. The Lattice simulation is  the unique
practical way to apply QCD in this energy region by so far.  A valence
lattice QCD result, where the pair creation through the Z graphs is
deleted in the connected insertions, supports the Goldstone boson
exchange picture~\cite{Liu:1998um,Liu:1999kq}, but Isgur pointed out
that this is unjustified considering that vQCD has a very different
spectrum from quench QCD and nature\cite{Isgur:1999ic}. In
Refs.~\cite{Okiharu2004,Okiharu2004a,Suganuma2011}, the confinement of
hadrons are investigated in SU(3) lattice QCD, and is well described
by the one-gluon-exchange (OGE) Coulomb plus string-theoretical linear
potential. However the spin-dependent part, that is, the hyperfine
interaction, which is more important for an constituent quark model,
is still introduced by hand.  In large $1/N_c$ approaches, Dan Pirjol
$et\ al.$ derived the correlations among the masses and mixing angles
and used it to test the different exchange models by study the
spin-flavor structure of the negative parity $L = 1$ excited
baryons\cite{Pirjol2009}.  They find that the experiment data
disfavour the pure gluon-exchange model. However as they indicted at a
footnote the $\Lambda(1405)$ considering there is a confusing state,
which may be not a pure three quark state, and the possible
long-distance contributions is also non-negligible. Hence their
conclusion is not reliable. In Ref~\cite{Pirjol:2010th} both gluon and
pion exchange can produce the observed data while the Isgur-karl model
need a much smaller exchanged vector meson in a range
$\mu\sim[0,400]$~MeV compared with the Lattice calculation of hybrid
meson masses and glueball $m_g\approx 800$~MeV. More precise use of
the Isgur-Karl model and its parameters should be improved to confirm
this conclusion.

Another way to evaluate the different hyperfine interactions is
applying the corresponding hyperfine interaction to the calculation of the
experimental observables and comparing with the experiment data
directly. By an effective matrix element method, Georgi $et\ al.$
analyze the P-wave baryon spectroscopy with both OGE and GBE hyperfine
interactions, and find
a smaller $\chi^2$ for the former in fitting the data. However the
authors claimed that it should not be interpreted as evidence in favor
of the chiral quark model picture~\cite{Collins:1998ny}.  A
comparative study has been done by calculating the effective
baryon-baryon interactions of the 64 lowest channels consisting of
octet and decuplet baryons with three types of hyperfine interaction, GBE,
OPE, and hybrid model, and find that these three models give similar
results, that is, all three models have reproduced the spectrum to
some extent~\cite{Wang:2002ha}.   As Isgur suggested the successful
description of the spectrum is only necessary but not sufficient to
determine whether a model is successful~\cite{Isgur2000a}.  It is
first applied by Chizma and Karl\cite{Chizma2003} though a simple
calculation of mixing angles of the low lying negative parity nucleon
resonances $S_{11}(1535)$ and $D_{13}(1650)$ and large deviation are
found. With those mixing angles the OPE can be excluded though the
calculation of the helicity amplitudes for the nucleon resonances
\cite{He2003}.  However with a remedy that the vector meson exchanged
are included the GBE model survived and  gave a reasonable mixing
angles for the low energy negative parity $N^*$~\cite{He2003a}.  Thus
a more comprehensive test of these two hyperfine interactions is essential.

Recently, the $\eta$ production processes, $\pi^{-}p\rightarrow\eta n$
and $\gamma p\rightarrow\eta p$, are investigated combinedly in a
chiral quark
approach~\cite{Li1995,zhao1998,Li1998a,Saghai2001a,Li1997f,He2008a,He2008}
equipped with OGE~\cite{He2009,He:2010ii}.  In this approach, the
observable are related with the different hyperfine interactions, 
such as OGE and GBE, through the configuration mixing of wave
functions. Hence it is a good place to check the exchange freedom in
the constituent quark models. The previous studies show a great
success of OGE mechanism. Hence in this work we will study
the effectiveness of GBE and OPE in the $\eta$ productions.

The structure of the paper is as follows. In the next section, a
sketch of the theoretical framework of our work will be
presented. In section~\ref{Sec:Res}, the numerical results  will be
reported and a discussion about the mixing angles of low lying
negative resonances will be presented also. Summary and conclusion
will be given in the last section.

\section{Theoretical Frame}\label{Sec:Theo}

In this section we recall briefly the chiral quark
approach and the three types of hyperfine interaction.
In the chiral quark model approach the amplitudes for a certain
resonance can be written as \cite{Li1997f,He2009,He:2010ii},
\begin{equation}\label{42}
{\cal M}_{N^*}=\frac {2M_{N^*}}{s-M_{N^*}^2-iM_{N^*}\Gamma({\bf q})}e^{-\frac {{\bf
k}^2+{\bf q}^2}{6\alpha^2}} {\cal O}_{N^*},
\end{equation}
where $\sqrt {s}$ is the total energy of the system, ${\bm
k}$ and ${\bm q}$ are the momenta of initial and final states in the
CM frame, and $M_{N^*}$ is the mass of the corresponding resonance and
$\alpha$ is the harmonic oscillator constant.
$\Gamma({\bf q})$ in Eq. (\ref{42}) is the total width of the
resonance, and a function of the final state momentum ${\bf q}$. The
${\cal O}_{N^*}$, the transition amplitude for pseudoscalar meson
production through photon and meson baryon scattering,  is
determined by the structure of each resonance, and takes,
respectively, the following CGLN form:
\begin{eqnarray}
\label{63} {\cal O}^\gamma_{N^*}&=&if_{1l\pm}  {\bf \sigma} \cdot {\bf
\epsilon}+ f_{2l\pm} {\bf \sigma} \cdot {\bf \hat{q}} {\bf \sigma}
\cdot ({\bf \hat{k}} \times {\bf \epsilon})+ if_{3l \pm} {\bf
\sigma} \cdot {\bf \hat{k}} {\bf \hat{q}} \cdot {\bf
\epsilon}\nonumber\\
&+&
if_{4l\pm}  {\bf \sigma} \cdot {\bf \hat{q}}{\bf \epsilon}\cdot {\bf
\hat{q}},\nonumber\\
{\cal O}^m_{N^*}&=&f_{1l\pm}+f_{2l\pm}{\bf
\sigma}\cdot\hat{\bf q}{\bf \sigma}\cdot\hat{\bf k}.
\end{eqnarray}

As we found in Ref.~\cite{He2008a}, with the helicity amplitudes of
photon transition and meson decay, we can directly obtain the CGLN
amplitudes for each resonance in terms of Legendre polynomials
derivatives.  We can connect the transition amplitudes with the
multipole coefficients as

\begin{eqnarray}
  f_{l\pm}&=& \mp A_{l\pm}= \frac{1}{2\pi(2J+1) }
[\frac{ E_{N_f}E_{N_i}}{
M^2_{N^*}}]^{1/2}A^{f}_{\lambda}A^{i}_{\lambda};,
\end{eqnarray}
where $J$ is the angular momentum of corresponding resonance and
$E_{N_f}$ and $E_{N_I}$ are the energies of incoming and final
nucleons.
The photoexcitation helicity amplitudes $A_\lambda^\gamma$, as well as
the strong decay amplitudes $A^m_\lambda$, is related to the matrix
elements of interaction Hamiltonian~\cite{Copley1969} as following,

\begin{eqnarray}A_\lambda  &=&\sqrt{\frac{2\pi}{k}}\langle
{N^*};J\lambda|H_{e}|N;\frac{1}{2}\lambda-1\rangle,  \\
A^m_\lambda&=&\langle N;\frac{1}{2}\nu|H_{m}|{N^*};J\lambda\rangle.
\end{eqnarray}
Here the $H_m$ is obtained from the effective chiral Lagrangian.

Except the transition Hamiltonian the wave function of the nucleon
resonances are essential to obtain the transition amplitude in quark
model, which is derived from the mass Hamiltonian in a
harmonic oscillator basis, $i.\ e.$ the $SU(6)$ wave functions.
Generally the Hamiltonian can be written as the following form

\begin{equation}
H=\sum_i \left( m_i+{{\bf p}_i^2 \over 2m_i}
\right)+\sum_{i<j}[V_{conf}(i,j)+V_{hyp}(i,j)].
\end{equation}
where ${\bf p}_i$ and $m_i$ are the momentum and the mass of $i$th quark.
$V_{conf}(i,j)$ is the confinement potential and $V_{hyp}(i,j)$
denotes the hyperfine interaction between $i$th and $j$th quark.

The OGE hyperfine interactions can be derived from the
one gluon exchange between two constituent quarks by non-relativization 
directly as given by Isgur
\cite{Isgur1978a,Isgur1979,Chizma2003},
\begin{eqnarray}
H^{ij}_{\rm hyp}&=&{2\alpha_s\over 3 m_i m_j} \left\{ {8\pi\over 3}
{\bf S}_i\cdot{\bf S}_j\delta^3({\bf r}_{ij}) +{1\over r_{ij}^3}
S_{ij}\right\} \label{Eq:OGE}
\end{eqnarray}
where  $\alpha_s$ is an effective quark-gluon fine-structure constant,
${\bf S}_i$ is the spin of $i$th quark, $S_{ij}=3{\bf S}_i\hat{\cdot{\bf r}}_{ij}{\bf
S}_j\cdot\hat{{\bf r}}_{ij}
- {\bf S}_i\cdot{\bf S}_j$ with ${\bf r}_{ij}$ is the separation
between $i$th and $j$th quark. For the confinement, we follow the
original paper by Isgur\cite{Isgur1979}.

The Hamiltonian of GBE model is more complicated. In this work we will
consider all pseudoscalar, vector and scalar meson exchanges except
the stange mesons $K^{(*)}$ which can not be exchnaged between $u$ or
$d$ quarks. The interactions
is also derived by the non-relativization  and written
as
\begin{eqnarray}
V_{hyp}(i,j)
&=&\sum^3_{a=1}\left[V_{\pi}(i,j)+V_{\rho}(i,j)\right]\lambda^a_i\cdot
\lambda^a_j\nonumber\\
&+&\left[V_{\eta}(i,j)+V_{\omega_8}(i,j)\right]
\lambda^8_i\cdot\lambda^8_j
\nonumber\\&+&
\frac{2}{3}\left[V_{\eta'}(i,j)+V_{\omega_0}(i,j)\right]+V_{\sigma}(i,j)
,
%+\nonumber\\[4pt]
%&&\right]
\label{Eq:GBE}
\end{eqnarray}
where $\lambda_i^a$ denote the Gell-Mann flavor matrices.
The explicit expressions of the different meson-exchanges
have been given explicitly in Ref.~\cite{He2003a,Wagenbrunn2000}.
In GBE model, a modified confinement potential is used as the form
$$V_{conf}(i,j)\ =\ V_0\ +\ C\ r_0\ (1\ -\ e^{-r_{ij}/r_0}\
).$$
The OPE model can be reached by turning off other exchanges except the
pion meson.

\section{Fitting procedure and numerical results}\label{Sec:Res}

In this work we following the fitting procedure in Ref.~\cite{He:2010ii}.
The data sets to be fitted
for differential cross-section and
 the polarized beam asymmetry of $\gamma p\rightarrow
\eta p$ and  differential cross-section of
$\pi^-p\rightarrow \eta n$ is listed in Table~\ref{Tab:data}.
\begin{table}[h!]
\caption{The data sets for the $\gamma p \to \eta p$
differential cross section (rows 2 to 7) and polarized beam asymmetry
(rows 8 and 9); differential cross section of the reaction $\pi^- p
\to \eta n$ (rows 10 to 14). The range of energy in the CM frame and
the number of data point are in the 3 and 4 column respectively.}
\renewcommand\tabcolsep{0.15cm}
\begin{tabular}{l|lrrrc}  \hline\hline
Observable & Collaboration/author  & W (GeV)          & $N_{dp}$    \\ \hline
$\frac{d\sigma}{d\Omega}~(\gamma p \to \eta p)$
           &MAMI94~\cite{Krusche1995} &   1.49 - 1.54  & 100  \\
           &CLAS09~\cite{Williams2009} &   1.68 - 2.80  &1081  \\
           &ELSA05~\cite{Crede2005} &   1.53 - 2.51 & 631  \\
           &ELSA09~\cite{Crede2009} &   1.59 - 2.37 & 680  \\
           &LNS06~\cite{Nakabayashi2006} &    1.49 - 1.74 & 180  \\
           &GRAAL07~\cite{Bartalini2007} &    1.49 - 1.91  & 487  \\\hline
$\Sigma~(\gamma p \to \eta p)$
           &ELSA07~\cite{Elsner2007} &    1.57 - 1.84 & 34  \\
           &GRAAL07~\cite{Bartalini2007} &    1.50 - 1.91  & 150  \\\hline
$\frac{d\sigma}{d\Omega}~(\pi^-p\rightarrow \eta n)$
           &Prakhov~{\it et al.}~\cite{Prakhov2005}&    1.49 - 1.52  & 84 \\
           &Deiinet~{\it et al.}~\cite{Deinet1969} &     1.51 - 1.70  & 80  \\
           &Richards~{\it et al.}~\cite{Richards1970} &     1.51 - 1.90  & 64  \\
           &Debenham~{\it et al.}~\cite{Debenham1975} &     1.49 - 1.67  & 24  \\
           &Brown~{\it et al.}~\cite{Brown1979} &     1.51 - 2.45  &102  \\
\hline \hline\end{tabular} \label{Tab:data}

\end{table}

In summary, 3697 experimental points will be used in the fitting
procedure.  Here the CLAS02~\cite{Dugger2002}is removed because it
is not consistent with the new data CLAS09 as shown in
Ref.~\cite{He:2010ii}.  In the previous work~\cite{He:2010ii}, the
spectrum, which is fitted to the values suggested by Particle Data
Group\cite{PDG}, is included only to give a constraint to the
parameters, so the $\chi^2$ for photon and $\pi$-induced productions
and spectrum are summed up directly. However for investigating the
hyperfine interaction of nucleon resonances, the spectrum is very
important. Hence a weight factor 150 are introduced for the $\chi^2$
of spectrum in this work. Besides, a weight factor 10 is also
introduced to $\pi$-induced production.

With the formalism in the previous section we calculate the
spectrum of nucleon resonances with mass $M<2$~GeV and the
$\Delta(1232)$ and observables for the $\eta$ productions. In the low
energy region the resonances in the Particle Data Group are adopted
besides a new $S_{11}$ states.  In the high energy region a
Reggeized treatment are applied. The theoretical results are fitting
to the experimental data by the MINUIT program. To do so, we have
a total of 25(23) free parameters for GBE (OPE) model and 19 free
parameters for OGE, which will be presented explicitly in the
following subsections.

\subsection{Results for the spectrum}

The mass spectrum of nucleon resonances have been studied in various
models. In the literatures, OGE, OPE and GBE models can reproduce the
mass spectrum successfully
~\cite{Isgur1978a,Isgur1979,Melde2007,Choi2010,Glozman1996}.
We have checked it with the interaction Hamiltonian given in
Eqs.(~\ref{Eq:OGE}) and (\ref{Eq:GBE}) by fitting the values of
Particle Data Group. All three models can give excellent description
of the mass spectrum with reasonable parameters as expected. However,
if the data of observable listed in Table~\ref{Tab:data} are included
in the fitting, the OPE can not give any acceptable results. Hence in
this subsection only the results of GBE and OGE models are presented.

We list the parameters related to the mass spectrum for GBE and OGE
models in Tables \ref{Tab:OGE} and \ref{Tab:GBE}.
\begin{table}[h!]
\caption{\label{Tab:OGE} Parameters of the  OGE model related to the
mass spectrum.}
\begin{center}
\renewcommand\tabcolsep{0.7cm}
\begin{tabular}{lcc}
\hline\hline  Parameter                 &
Ref.\cite{He:2010ii}& current work   \\ \hline
 $m_q$[MeV]                        &  312                & 310    \\
 $\alpha_{ho}$[MeV]                &  348                & 309    \\
$\alpha_s$                   &  1.96               &  1.61   \\
   $\Omega$[MeV]                     &  437                & 428   \\
   $\Delta$[MeV]                     &  460                & 456   \\ \hline
 $\chi^2_{spec}$               & 3.4   &1.3    \\
 \hline\hline
\end{tabular}
\label{Tab:parameter}
\end{center}
\end{table}

For the OGE model, the average $\chi^2_{spec}$ for the mass spectrum
decrease obviously, from 3.4 to 1.3, because the contribution of mass
spectrum in the total $\chi^2$ is weighted by a factor 150. The
parameters related to the spectrum have slight variation and are in the
reasonable ranges.

\begin{table}[h!]
\caption{\label{Tab:GBE} Parameters of the GBE model related to the
mass spectrum}
\begin{center}
\renewcommand\tabcolsep{0.4cm}
\begin{tabular}{lcc}
\hline\hline  Parameter                 &
Ref.\cite{Wagenbrunn2000}
& current work   \\ \hline
 $m_q$ [MeV]           & 340    & 251     \\
  $\alpha_{ho}$[MeV]                &  $--$                & 316   \\
 ${(g^V_{v,8}+g^T_{v,8})}^2$/4$\pi$                & 1.31   &1.39    \\
 $g^2_{ps}$/4$\pi$                                 & 0.67    & 0.80   \\
 $\Lambda_\pi$ [fm$^{-1}$]  & 0.7    & 0.45    \\
 $\Lambda_\rho$ [fm$^{-1}$] & 1.2    & 1.58  \\
 $\kappa            $      & 1.2    &1.83   \\
 $C$ [fm$^{-2}$]           & 2.53   & 2.13   \\
 $r_0$ [fm]                & 7      & 6       \\\hline
 $\chi^2_{spec}$               & $--$   & 1.6    \\
\hline\hline

\end{tabular}
\end{center}
\end{table}

For GBE model, we compare our parameters with the ones in Ref.~\cite{Wagenbrunn2000}
where only the spectrum is considered. As indicated in
Table~\ref{Tab:GBE}, the non-strange quark mass ($m_q$) is smaller
compared with the one in the original paper
but still in the reasonable range in constituent quark model,
200$\sim$350MeV. In order to prevent a proliferation
of free parameters, a universal coupling constant $g^2_{ps}$/4$\pi$ is
assumed to be equal to the quark-pion coupling constant
$g^2_{\pi}$/4$\pi$, which is popular used in the literatures
~\cite{Zhang1997,Huang2003}. Such treatment is also applied to the vector meson
case.  The fitted parameters are consistent with the results by Graz
group~\cite{Wagenbrunn2000}.  The confinement
strength, $C\approx2.15~fm^{-2}$, is close to the Lattice one
\cite{Creutz}.  The average
$\chi^2_{spec}$ in GBE model is about 1.6, which is a little larger
than the OGE model.

\begin{figure}[h!]
\includegraphics[bb=20 300 490 740 ,scale=0.5,clip]{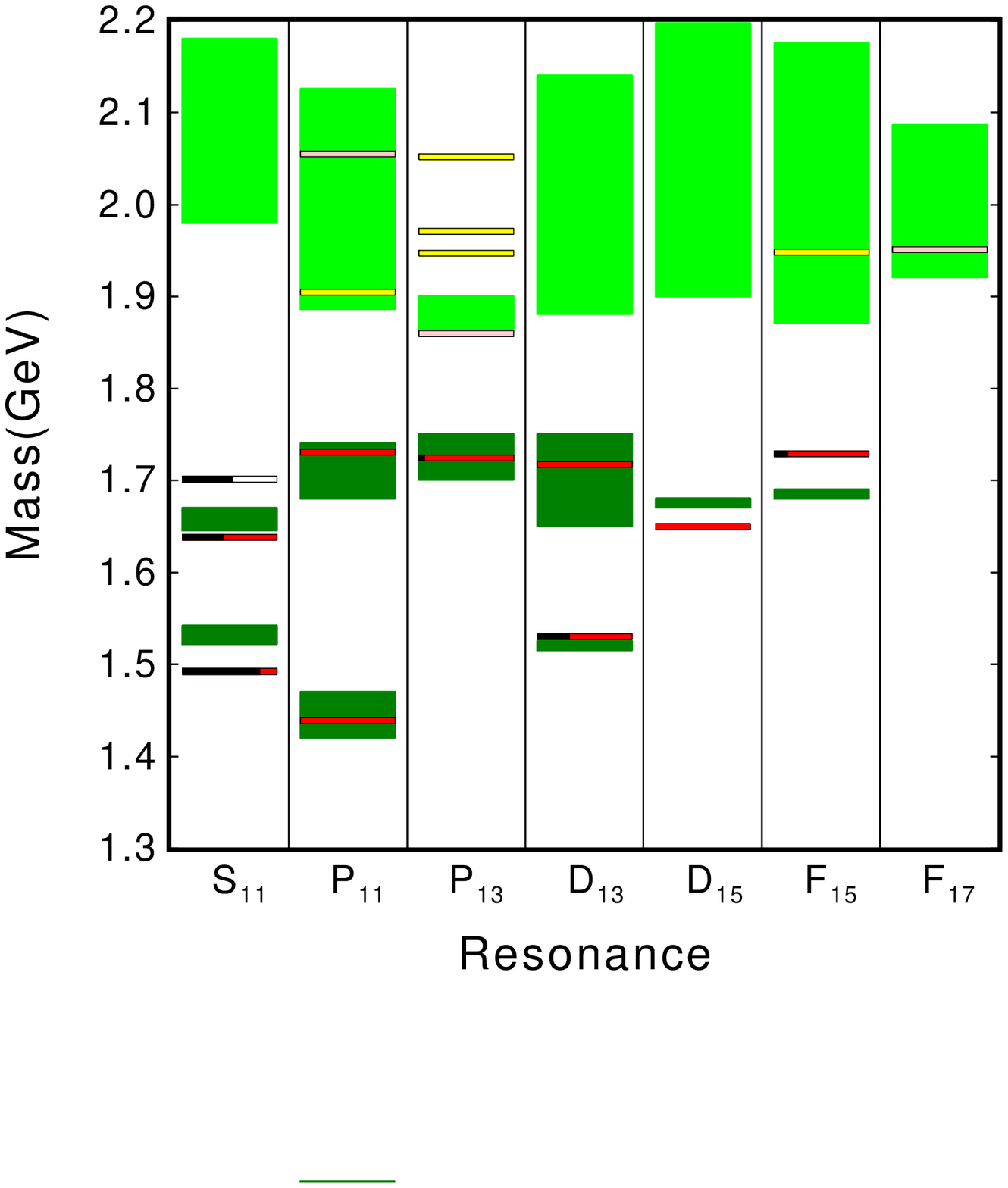}
\includegraphics[bb=20 300 490 740 ,scale=0.5,clip]{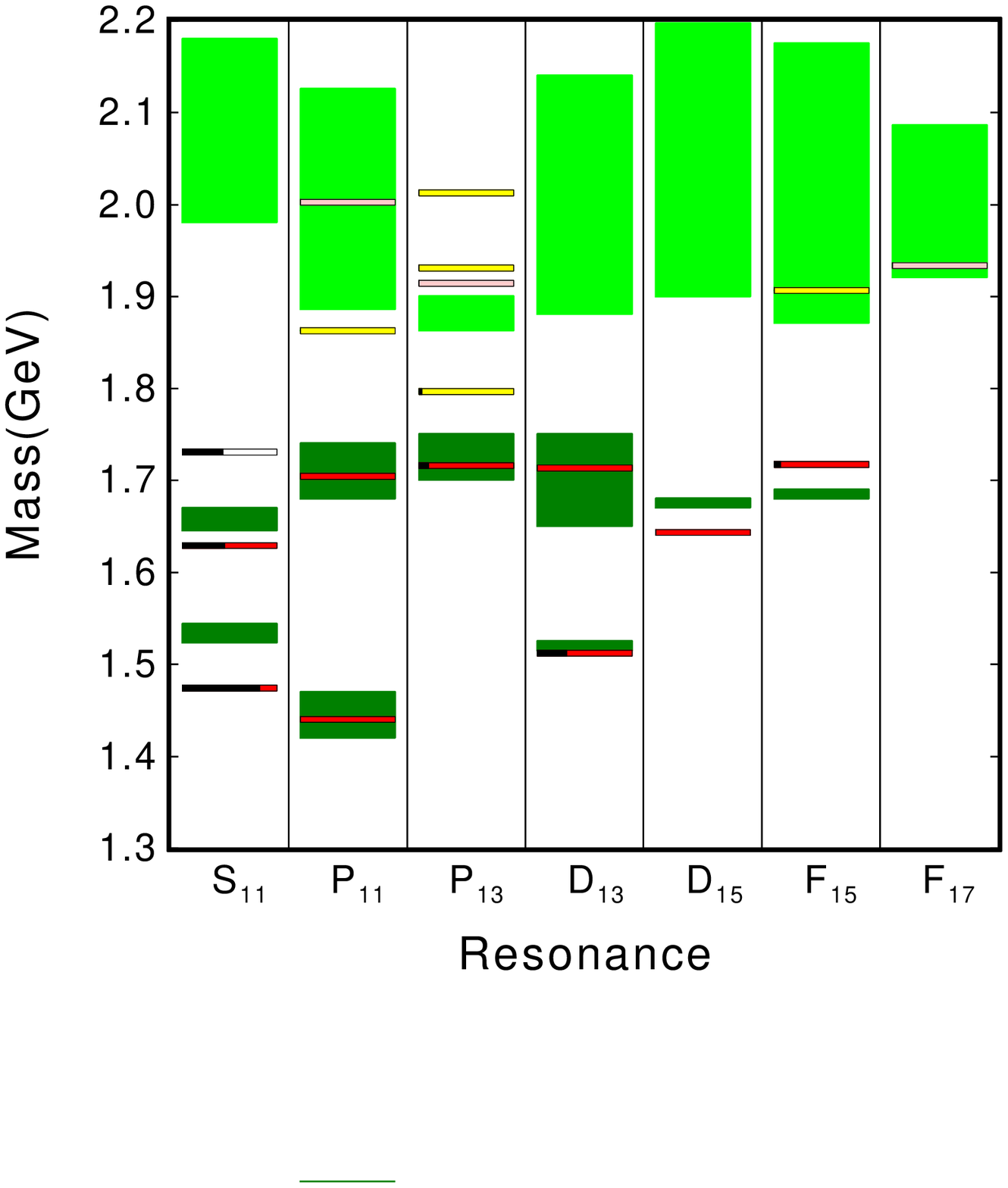}
\caption{{\footnotesize The spectrum of baryon
resonances from PDG~\cite{PDG} (*,** light green bands,***,**** dark
green bands) and from the
current work for known (*,** light red bars,***,**** dark
red bars), missing (yellow bars), and new
(white bar) resonances.  The black bars indicate $\log\Delta\chi^2/\chi^2_0$, the logarithms of variations of
$\chi^2$ after turning off the corresponding resonance compared the
one of full
model $\chi^2_0$. The higher one is for the OGE model and the lower one
is for the GBE model.}
\label{Fig:spcl}}
\end{figure}

With the above parameters the theoretical results of the mass spectrum
by OGE and GBE models are presented in Fig.~\ref{Fig:spcl} and compared
with the values in Particle Data Group.

Compared with the original papers~\cite{Isgur1978a,Isgur1979}, the
results of the OGE model in the current work have an excellent agreement with
the values of Particle Data group due to the parameters obtained by
fitting not chosen by hand as in the original paper. For the GBE
model, most states are well reproduced as the original work. However
through the relative ordering of the lowest positive- and
negative-parity excitations is obtained rightly, the gap is smaller.
If we fitting the mass spectrum only, the gap can be well reproduced.
It supports Isgur's suggestion  that the mass spectrum is not enough to
judge an model.
Generally the two models give very similar description of the mass
spectrum and reproduce the values suggested by Particle Data Group.
In the OGE model the two star $P_{13}(1900)$ in Particle Data Group can be assigned
as the second $P_{13}$ states while in GBE the mass of second states is too
low, so we assign the $P_{13}(1900)$ as the third one. Because it is
not important in the channels considered in this work we will not
give any conclusion about this state here.

%%%%%%%%%%%%%%%%%%%%%%%%%%%%%%%%%%%%%%%%%%%%%%%%%%%%%%%%%%%%%%%%%%%%%%%%%%

\subsection{Results  for  $\eta$ productions}

In Table~\ref{Tab: Para}, the remaining parameters in GBE and OGE
models, which are only involved in the $\eta$ productions, are listed.
\begin{table}[ht!]
\caption{{\footnotesize Parameters related to the $\eta$ productions, where
$M$ and $\Gamma$ are in
MeV.}}
\renewcommand\tabcolsep{0.25cm}
\begin{tabular}{ll|ccc}  \hline\hline
    & Parameter                    &  GBE & OGE
    &Ref.~\cite{He:2010ii}     \\ \hline
  & $g_{\eta NN}$                &  0.146              & 0.290 & 0.276 \\ \hline
$P_{13}(1720)$:
  & $C^\gamma_{P_{13}(1720)}$    &  0.107               & 0.172 & 0.22 \\
  & $C^\pi_{P_{13}(1720)}$       & -0.916              & -0.823 &  -0.85\\ \hline
New $S_{11}$: &$M$        & 1730                & 1701 &1700 \\
              &$\Gamma$   & 319                 & 469  &473\\
              &$C_{N^*}$  & 0.509                & 1.11 &1.18  \\ \hline
$N(1535)$:    &$M$        & 1534                & 1532 &1532  \\
              &$\Gamma$   & 169                & 144  &140 \\\hline
$u$-channel:
  & $C^\gamma_{u}$               & 0.36                & 0.63 &0.71  \\
  & $C^\pi_{u}$                  & 1.53                & 1.26 &1.39  \\ \hline
{\it t}-channel   &$g_{\rho qq}$     & 1.43            & 1.99 &1.90\\
          &$\kappa_{\rho qq}$     & -0.14          &-0.25 &-0.20 \\
          &$g_{\omega qq}$   & 3.33                & 4.87 &4.88 \\
          &$\kappa_{\omega qq}$   & -0.15          & -0.30&-0.26 \\\hline
$\chi^2$      &$\gamma p\to\eta p$   &       2.7       &  2.4  &2.5
\\
             &$\pi^- p\to\eta n$   &   1.6             &  1.3  &1.3
\\
\hline \hline\end{tabular} \label{Tab: Para}
\end{table}

For the OGE model after weighted the contribution of spectrum, the
parameters is almost unchanged. All parameters for the GBE model are
consistent with the ones for the OGE model in the reasonable ranges. The
coupling constant $g_{\eta NN}$ is $0.146$ in GBE model, which is
smaller than the one in OGE.  Comparable values are also reported in
Refs.~\cite{Tiator1994,Kirchbach1996,Zhu2000,Stoks1999}. Here the
strengths for $P_{13}(1720)$ and $u$-channel are consistent in the two
models. The mass and decay width of $S_{11}(1535)$ are fitted
directly due to that it can not be produced by the current mass Hamiltonian
and is important for the $\eta$ productions. The theoretical values are well 
comparable to the suggest values of Particle Data Group. An additional
$S_{11}$ with a mass about 1730MeV is introduced for the GBE model as well as the OGE model
to reproduce the observables for $\eta$ productions. Though
average $\chi^2$ for spectrum in the OGE model decrease from 3.4 to 1.3 as shown in the
previous subsections, the average $\chi^2$ for $\eta$ photoproduction
increase a little. The average $\chi^2$s in the GBE model for two
process are acceptable though a little large than the ones in GBE
model. 

In Fig.~\ref{Fig:spcl}, the variations of $\chi^2$ after turning off
the corresponding resonance contribution, without further
minimizations, are presented with the mass spectrum. As the mass
spectrum,  the GBE and OGE models lead to  similar mechanism  
of the two channels considered in this work. Among the eighty
resonances considered, six resonances, three $S_{11}$ states, first $P_{13}$
states, the first $D_{13}$ state and $F_{15}$, indicate their
existence in the channels considered in this work. The contributions
from all five so-called ``missing resonances'' is negligible in both
GBE and OGE models.

\begin{figure}[h!]
  \includegraphics[ bb=60 380 550 740 ,scale=0.53 ]{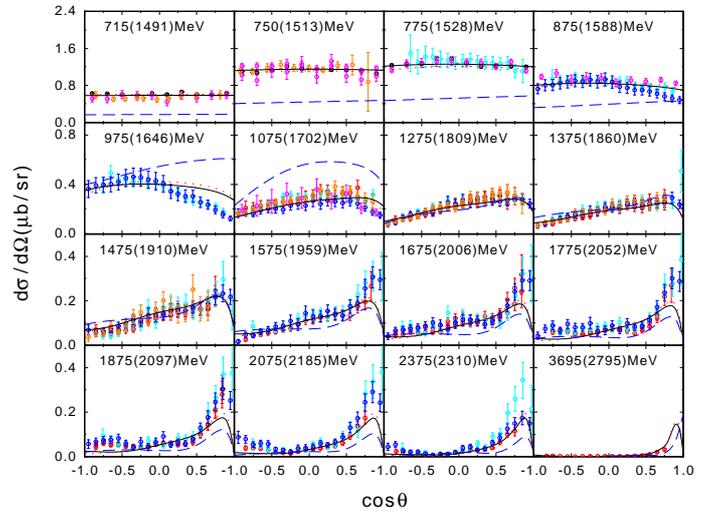}
\caption{{\footnotesize Differential cross section for $\gamma p \to \eta p$ as
a function of $cos \theta_\eta$ for various values of photon energy in the lab frame.
The values in parenthesis are the corresponding total energy
of the system $W$.
The curves are: GBE (full),
OGE (dotted), and OPE (Dashed).
Data are from Refs.\cite{Krusche1995,Williams2009,Crede2005,Crede2009,Nakabayashi2006,Bartalini2007,Dugger2002}.}
\label{Fig:gpep}}
\end{figure}

\begin{figure}[h!]
  \includegraphics[ bb=60 530 550 740 ,scale=0.53 ]{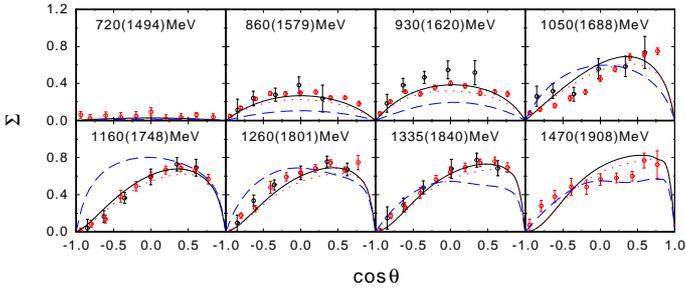}
\caption{{\footnotesize Same as Fig. \ref{Fig:gpep}, but for polarized
beam asymmetry for $\vec{\gamma} p\rightarrow \eta p$.
Data are from Refs.~\cite{Bartalini2007,Elsner2007}.}
\label{Fig:gpepBA}}
\end{figure}

The explicit results for the differential cross section and polarized beam asymmetry for
$\gamma p\rightarrow \eta p$ at some energy points are presented in the
Figs.~\ref{Fig:gpep} and~\ref{Fig:gpepBA}.

The differential cross sections in OGE and GBE models are almost same
as each other and have agreement with the experiments. The results for
OPE model are presented in the same figure also, and one can find the
experimental data can not be reproduced especially in low energy region,
which is from
the wrong mixing angles, and will be discussed later. For the higher
energy region, where the $t$-channel contribution is dominant, the
failure of OPE model to fit the data is not from the t-channel itself
but the wrong description of resonances in low
energy region, that is, the large discrepancy in the low energy region make it
difficult to find a set of parameters to give acceptable results in the
whole energy region.

For the polarized beam asymmetries, the curves for the OGE and GBE
model have a slight discrepancy but both agree with the data. The OPE
model can not reproduce the data as shown in results of the differential cross
section.

\begin{figure}[ht!]
  \includegraphics[ bb=60 370 550 740 ,scale=0.53 ]{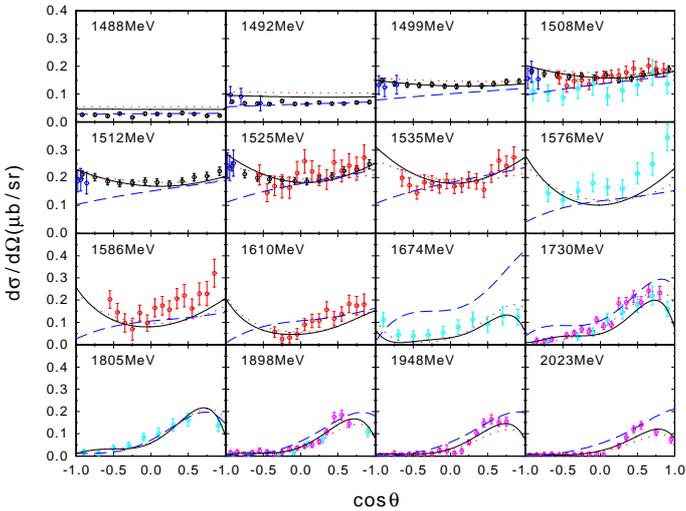}
\caption{{\footnotesize Differential cross section for $\pi^- p \to \eta n$ as a
function of $cos {\theta _\eta}$ for various values of the total energy
of the system $W$.
The curves are: GBE (full),
OGE (dotted), and OPE (Dashed).
Data are from  Refs.~\cite{Prakhov2005,Deinet1969,Richards1970,Debenham1975,Brown1979}.}
\label{Fig:ppen}}
\end{figure}

The differential cross sections for the pion induced $\eta$ production
are presented in the Fig.~\ref{Fig:ppen}.
As in the photoproduction case, the full model with OGE and GBE model
reproduce the data in a very similar ways specially near the
threshold while the OPE
model can not reproduce the experimental data .

\subsection{The mixing angles of negative parity resonances}

As suggested and applied by many authors
\cite{Chizma2003,Isgur2000a,Saghai2010}, the mixing angle of
two pairs of the negative parity states, $\theta_S$ for $S_{11}(1535)$
and $S_{11}(1650)$ and $\theta_D$ for $D_{13}(1520)$ and
$D_{13}(1700)$, is very important to understand the structure of
resonances and underlying dynamics of the interquark interaction. In
mid-seventies of the past century, Hey $et\ al.$ \cite{Hey1975} have
determined the mixing angles from the decays of baryon resonances
as $\theta_S=-31.9^\circ$ and $\theta_D=10.4^\circ$.
The standard values of mixing angles in OGE model is
$\theta_S=-32^\circ$ and $\theta_D=6^\circ$, which can be obtained without
parameter-dependent form the OGE hyperfine interaction\cite{Isgur1978a}.
Recently Dan Pirjol et al. \cite{Pirjol2009} apply the
large-$N_{c}$ expansion approach to the determination of the mixing
angles within the following ranges: $0^\circ\leq\theta_S\leq35^\circ$
and $0^\circ\leq\theta_S\leq45^\circ$.

The OPE model in the literature gives the values of mixing angles as
$\theta_S=26^\circ$ and $\theta_D=-53^\circ$~\cite{Chizma2003}, which
can not reproduce the helicity amplitudes~\cite{He2003}.  Glozman
\cite{Glozman1999} pointed out that in the $\rho$-exchange tensor
interaction dominate over the $\pi$-exchange tensor interaction in the
P-wave baryons. The $\rho$-like exchange contribute a positive sign to
the mixing angle while the $\pi$-like exchange contribute a negative
sign in the reasonable parameter range. The mixing angles of $S_{11}$
and $D_{13}$ obtained here in the GBE model are $31^{\circ}$ and
$-6^{\circ}$, respectively, which is almost same as values of the OGE
model.  Combined with the agreement with the experiments of the
theoretical results for $\eta$ productions, it suggests the data for
the $\eta$ productions support these values strongly. Here the OPE
models give $24^\circ$ and $-47^\circ$, which is close to the ones in
Ref~\cite{Chizma2003,He2003a}, $25.5^\circ$ and $-52.5^\circ$, 
obtained from the mass spectrum directly. As shown in the above
subsections, these values will lead to wrong differential cross
sections.

\section{Summary and conclusion}\label{Sec:Conclu}

In this work the spectrum and differential cross sections and
polarized asymmetries are calculated with the GBE, OGE and OPE
hyperfine interactions.
The GBE model shows a similar pattern to the OGE in spectrum,
observables, even the mechanism of the reactions while the OPE can
be excluded by the large discrepancy of the observables for the
$\eta$ productions.

The mixing angles of negative parity states are almost same for GBE
and OGE models, which indicates the experimental data of $\eta$
productions support the convention mixing angles
$\theta_S=-32^\circ$ and $\theta_D=6^\circ$. Since both GBE and OGE
can produce the right mixing angles for the four negative parity
resonances, which play dominant roles in the $\eta$ production, the
$\eta$ production is not a right place to check these two models.
From the theoretical point of view, it is necessary to perform more
investigations into CQMs through other channels , such as $\gamma
p\rightarrow K \Lambda$ and $\pi^-p\rightarrow K \Lambda$. In these
reaction channels, the main contributions are not only from negative
parity states $S_{11}(1535)$ and $S_{11}(1650)$ but also from
several positive parity states $P_{11}(1700)$ and $P_{13}(1720)$ and
so on. The predictions for the wave function structure of positive
parity states in CQMs will provide further information for
distinguishing them. Many investigations \cite{sara,nik} have shown
that in strangeness production channel, missing resonances play
essential role in reproducing the rich single and double
polarization data, which may be a   
way to investigate the hyperfine interaction between the constituent
quarks.
%------------------------------------------
%

\section*{Acknowledgements}

This project is supported by the National Natural Science Foundation
of China under Grants No. 10905077, No.  11035006, the Ministry of
Education of China (the project sponsored by SRF for ROCS, SEM under
Grant No. HGJO90402) and Chinese Academy of Sciences (the Special
Foundation of President under Grant No. YZ080425).

%\bibliography{/home/junhe/Paper/bibliography}

\begin{thebibliography}{62}
%\cite{Bowman:2005vx}
\bibitem{Bowman:2005vx}
  P.~O.~Bowman, $et\ al.$,
  %``Unquenched quark propagator in Landau gauge,''
  Phys.\ Rev.\  D { 71}(2005) 054507.
  %%CITATION = PHRVA,D71,054507;%%



%\cite{Melikhov:2004uk}
\bibitem{Melikhov:2004uk}
  D.~Melikhov and S.~Simula,
  %``A correspondence between QCD sum rules and constituent quark models,''
  Eur.\ Phys.\ J.\  C { 37} (2004) 437.
  %%CITATION = EPHJA,C37,437;%%
%Type = Article
\bibitem{Pirjol2009}
D.~Pirjol, C.~Schat, Phys. Rev. Lett.102 (2009) 152002.

%\cite{Pirjol:2010th}
\bibitem{Pirjol:2010th}
  D.~Pirjol and C.~Schat,
  %``1/Nc expansion and the spin-flavor structure of the quark interaction in
  %the constituent quark model,''
  Phys.\ Rev.\  D { 82} (2010) 114005
  %%CITATION = PHRVA,D82,114005;%%

%Type = Article
\bibitem{Isgur1978a}
N.~Isgur, G.~Karl, Phys. Rev.D18(1978)4187.
%Type = Article

\bibitem{Isgur1979}
N.~Isgur, G.~Karl, Phys. Rev.D19(1979)2653.
%Type = Article
\bibitem{Capstick1986}
S.~Capstick, N.~Isgur, Phys. Rev.D34(1986) 2809.
%Type = Article
\bibitem{Koniuk1980}
R.~Koniuk, N.~Isgur, Phys. Rev.D21(1980)1868.
%Type = Article
\bibitem{Capstick2000a}
S.~Capstick, W.~Roberts, Prog. Part. Nucl. Phys. 45(2000) S241--S331.
%Type = Article
\bibitem{Melde2007}
T.~Melde, W.~Plessas, B.~Sengl, Phys. Rev. C76(2007) 025204.
%Type = Article
\bibitem{Choi2010}
K.-S. Choi,W.~Plessas,R.~F. Wagenbrunn, Phys. Rev. C81(2010) 028201.
%\cite{Manohar:1983md}
\bibitem{Glozman1996}
L.~Y. Glozman,D.~O. Riska, Phys. Rep. 268(1996)263.
%Type = Article
\bibitem{Manohar:1983md}
  A.~Manohar and H.~Georgi,
  %``Chiral Quarks And The Nonrelativistic Quark Model,''
  Nucl.\ Phys.\  B { 234} (1984) 189.
  %%CITATION = NUPHA,B234,189;%%

%\cite{Collins:1998ny}
\bibitem{Collins:1998ny}
  H.~Collins and H.~Georgi,
  %``S(3) and the L = 1 baryons in the quark model and the chiral quark
  %model,''
  Phys.\ Rev.\  D { 59} (1999) 094010.
  %%CITATION = PHRVA,D59,094010;%%

%\cite{Liu:1998um}
\bibitem{Liu:1998um}
  K.~F.~Liu, $et\ al.$,
  %``Valence {QCD}: Connecting {QCD} to the quark model,''
  Phys.\ Rev.\  D { 59} (1999) 112001.
  %%CITATION = PHRVA,D59,112001;%%

%\cite{Liu:1999kq}
\bibitem{Liu:1999kq}
  K.~F.~Liu, $et\ al.$,
  %``Reply to Isgur's comments on valence QCD,''
  Phys.\ Rev.\  D { 61} (2000) 118502.
  %%CITATION = PHRVA,D61,118502;%%

%\cite{Isgur:1999ic}
\bibitem{Isgur:1999ic}
  N.~Isgur,
  %``Comment on 'Valence QCD: Connecting QCD to the quark model',''
  Phys.\ Rev.\  D { 61} (2000) 118501.
  %%CITATION = PHRVA,D61,118501;%%

%\cite{Okiharu:2004ve}
\bibitem{Okiharu2004}
  F.~Okiharu, H.~Suganuma and T.~T.~Takahashi,
  %``The tetraquark potential and flip-flop in SU(3) lattice QCD,''
  Phys.\ Rev.\  D {\bf 72} (2005) 014505.
  %%CITATION = PHRVA,D72,014505;%%

%\cite{Okiharu:2004wy}
\bibitem{Okiharu2004a}
  F.~Okiharu, H.~Suganuma and T.~T.~Takahashi,
  %``First study for the pentaquark potential in SU(3) lattice QCD,''
  Phys.\ Rev.\ Lett.\  {\bf 94} (2005) 192001.
  %%CITATION = PRLTA,94,192001;%%

%\cite{Suganuma:2011ci}
\bibitem{Suganuma2011}
  H.~Suganuma, T.~Iritani, F.~Okiharu, T.~T.~Takahashi and A.~Yamamoto,
  %``Lattice QCD Study for Confinement in Hadrons,''
  arXiv:1103.4015 [hep-lat].
  %%CITATION = ARXIV:1103.4015;%%





%\cite{Wang:2002ha}
\bibitem{Wang:2002ha}
  F.~Wang, J.~L.~Ping, H.~R.~Pang and J.~T.~Goldman,
  %``Which Constituent Quark Model Is Better?,''
  Mod.\ Phys.\ Lett.\  A { 18} (2003) 356.
  %%CITATION = MPLAE,A18,356;%%
%Type = Article
\bibitem{Isgur2000a}
{N.~Isgur},
 {Phys. Rev.} {D62}
  ({2000}) {054026}.


%Type = Article
\bibitem{Chizma2003}
J.~Chizma,G.~Karl, Phys. Rev.D68(2003)054007.

%\cite{Jun:2003cu}
\bibitem{He2003}
  J. He and Y. B. Dong,
  %``Test of one-pion exchange and one-gluon exchange through mixing angles of
  %negative parity N* resonances in electromagnetic transitions,''
  Phys.\ Rev.\  D {\bf 68} (2003) 017502.
  %%CITATION = PHRVA,D68,017502;%%



%Type = Article
\bibitem{He2003a}
{J.~He}, {Y.-B. Dong},
 {Nucl. Phys.} {A725}
  ({2003}) {201--210}.
%Type = Article
\bibitem{Li1995}
{Z.-P. Li},
 {Phys. Rev.} {C52}
  ({1995}) {1648--1661}.
%Type = Article
\bibitem{zhao1998}
{Q.~Zhao}, {Z.-p. Li},
  {C.~Bennhold},
 {Phys. Rev.} {C58}
  ({1998}) {2393--2413}.
%Type = Article
\bibitem{Li1998a}
{Z.-p. Li}, {B.~Saghai},
 {Nucl. Phys.} {A644}
  ({1998}) {345--364}.
%Type = Article
\bibitem{Saghai2001a}
{B.~Saghai}, {Z.-p. Li},
 {Eur. Phys. J.} {A11}
  ({2001}) {217--230}.
%Type = Article
\bibitem{Li1997f}
{Z.-P. Li}, {J.~Bao},
 {Europhys. Lett.} {39}
  ({1997}) {599--604}.
%Type = Article
\bibitem{He2008a}
{J.~He}, {B.~Saghai}, {Z.~Li},
 {Phys. Rev.} {C78}
  ({2008}) {035204}.
%Type = Article
\bibitem{He2008}
{J.~He}, {B.~Saghai}, {Z.~Li},
  {Q.~Zhao}, {J.~Durand},
 {Eur. Phys. J.} {A35}
  ({2008}) {321--324}.

 %Type = Article
\bibitem{He2009}
{J.~He}, {B.~Saghai},
 {Phys. Rev.} {C80}
  ({2009}) {015207}.

 %\cite{He:2010ii}
\bibitem{He:2010ii}
  J.~He and B.~Saghai,
  %``$\eta$ production off the proton in a Regge-plus-chiral quark approach,''
  Phys.\ Rev.\  C { 82} (2010) 035206.
  %%CITATION = PHRVA,C82,035206;%%



%Type = Article
\bibitem{Copley1969}
{L.~A. Copley}, {G.~Karl},
  {E.~Obryk},
 {Nucl. Phys.} {B13}
  ({1969}) {303--319}.
%Type = Article
\bibitem{Wagenbrunn2000}
{R.~F. Wagenbrunn}, {L.~Y. Glozman},
  {W.~Plessas}, {K.~Varga},
 {Nucl. Phys.} {A666}
  ({2000}).
%Type = Article
\bibitem{Krusche1995}
{B.~Krusche}, et~al.,
 {Phys. Rev. Lett.} {74}
  ({1995}) {3736--3739}.


%Type = Article
\bibitem{Williams2009}
{M.~Williams}, et~al.,
 {Phys. Rev.} {C80}
  ({2009}) {045213}.
%Type = Article
\bibitem{Crede2005}
{V.~Crede}, et~al.,
 {Phys. Rev. Lett.} {94}
  ({2005}) {012004}.
%Type = Article
\bibitem{Crede2009}
{V.~Crede}, et~al.,
 {Phys. Rev.} {C80}
  ({2009}) {055202}.
  %Type = Article
\bibitem{Nakabayashi2006}
{T.~Nakabayashi}, et~al.,
 {Phys. Rev. C} {74}
  ({2006}) {035202}.


%Type = Article
\bibitem{Bartalini2007}
{O.~Bartalini}, et~al.,
 {Eur. Phys. J.} {A33}
  ({2007}) {169--184}.
%Type = Article
\bibitem{Elsner2007}
{D.~Elsner}, et~al.,
 {Eur. Phys. J.} {A33}
  ({2007}) {147--155}.
%Type = Article
\bibitem{Prakhov2005}
{S.~Prakhov}, et~al.,
 {Phys. Rev.} {C72}
  ({2005}) {015203}.


  %Type = Article
\bibitem{Deinet1969}
{W.~Deinet}, et~al.,
 {Nucl. Phys.} {B11}
  ({1969}) {495--504}.
%Type = Article
\bibitem{Richards1970}
{W.~B. Richards}, et~al.,
 {Phys. Rev. D} {1}
  ({1970}) {10--19}.
%Type = Article
\bibitem{Debenham1975}
{N.~C. Debenham}, et~al.,
 {Phys. Rev.} {D12}
  ({1975}) {2545--2556}.


 %Type = Article
\bibitem{Brown1979}
{R.~M. Brown}, et~al.,
 {Nucl. Phys.} {B153}
  ({1979}) {89--111}.

 %Type = Article
\bibitem{Dugger2002}
{M.~B. Dugger}, et~al.,
 {Phys. Rev. Lett.} {89}
 ({2002}) {222002}.

 %Type = Article
 \bibitem{PDG}
K. Nakamura et al. (Particle Data Group), J. Phys. G {\bf37} (2010) 075021.

%\cite{Zhang:1997ny}
\bibitem{Zhang1997}
  Z.~Y.~Zhang, Y.~W.~Yu, P.~N.~Shen, L.~R.~Dai, A.~Faessler and U.~Straub,
  %``Hyperon nucleon interactions in a chiral SU(3) quark model,''
  Nucl.\ Phys.\  A {\bf 625} (1997) 59.
  %%CITATION = NUPHA,A625,59;%%

%\cite{Huang:2003we}
\bibitem{Huang2003}
  F.~Huang, Z.~Y.~Zhang, Y.~W.~Yu and B.~S.~Zou,
  %``A Study of pentaquark Theta state in the chiral SU(3) quark model,''
  Phys.\ Lett.\  B {\bf 586}, 69 (2004)
  [arXiv:hep-ph/0310040].
  %%CITATION = PHLTA,B586,69;%%
%Type = Article
\bibitem{Creutz}
M.~Creutz, {\it Monographs On Mathematical Physics},
Cambridge University Press, Cambridge(UK).

%Type = Article
\bibitem{Tiator1994}
{L.~Tiator}, {C.~Bennhold},
  {S.~S. Kamalov},
 {Nucl. Phys.} {A580}
  ({1994}) {455--474}.
%Type = Article
\bibitem{Kirchbach1996}
{M.~Kirchbach}, {L.~Tiator},
 {Nucl. Phys.} {A604}
  ({1996}) {385--394}.
%Type = Article
\bibitem{Zhu2000}
{S.-L. Zhu},
 {Phys. Rev.} {C61}
  ({2000}) {065205}.
%Type = Article
\bibitem{Stoks1999}
{V.~G.~J. Stoks}, {T.~A. Rijken},
 {Phys. Rev.} {C59}
  ({1999}) {3009--3020}.
%Type = Article
\bibitem{Saghai2010}
{B.~Saghai}, {Z.~Li},
 {Few Body Syst.} {47}
  ({2010}) {105--115}.
%Type = Article
\bibitem{Hey1975}
{A.~J.~G. Hey}, {P.~J. Litchfield},
  {R.~J. Cashmore},
 {Nucl. Phys.} {B95}
  ({1975}) {516}.
%Type = Techreport
\bibitem{Glozman1999}
{L.~Y. Glozman}, {arXiv:}
  {nul-th/9909021}.


\bibitem{sara}  A.V. Sarantseva,  V.A. Nikonov, A.V. Anisovich, E.
	Klempt, and U. Thoma,
 Eur.~Phys.~J.~A25(2005)441-453.

\bibitem{nik} V.A. Nikonov, A.V. Anisovich, E. Klempt,  A.V.
	Sarantseva, and U. Thoma,
 Phys.~Lett.~B662(2008)245.


\end{thebibliography}

\end{document}